# From Priors to Predictions: Explaining and Visualizing Human Reasoning in a Graph Neural Network Framework

Quan Do[1,3], Caroline Ahn[1,3], Leah Bakst[2,4], Michael Pascale[2,4], Joseph T. McGuire[1,2,3,4], Chantal E. Stern[1,2,3,4], Michael E. Hasselmo[1,2,3,4]

1. Graduate Program for Neuroscience, Boston University, Boston, MA
2. Department of Psychological and Brain Sciences, Boston University, Boston, MA
3. Center for Systems Neuroscience, Boston University, Boston, MA
4. Cognitive Neuroimaging Center, Boston University, Boston, MA

## Abstract

Humans excel at solving novel reasoning problems from minimal exposure, guided by inductive biases—assumptions about which entities and relationships matter. Yet the computational form of these biases and their neural implementation remain poorly understood. We introduce a framework that combines Graph Theory and Graph Neural Networks (GNNs) to formalize inductive biases as explicit, manipulable priors over structure and abstraction. Using a human behavioral dataset adapted from the Abstraction and Reasoning Corpus (ARC), we show that differences in graph-based priors can explain individual differences in human solutions. Our method includes an optimization pipeline that searches over graph configurations—varying edge connectivity and node abstraction—and a visualization approach that identifies the computational graph, the subset of nodes and edges most critical to a model's prediction. Systematic ablation reveals how generalization depends on specific prior structures and internal processing, exposing why human-like errors emerge from incorrect or incomplete priors. This work provides a principled, interpretable framework for modeling the representational assumptions and computational dynamics underlying generalization, offering new insights into human reasoning and a foundation for more human-aligned AI systems.

## Introduction

When confronted with a new challenge in an unfamiliar and puzzling situation, humans can rapidly formulate a hypothesis based on limited interactions and come up with a solution tailored to the specific problem. This reasoning capability consistently appears in a broad spectrum of human endeavors, ranging from mathematics, science, and technology to social, cultural, and political engagements, and remains a hallmark of human intelligence, yet to be rivaled by state-of-the-art AI (Artificial Intelligence). However, coming up with a quick solution does not guarantee a correct response. It is therefore beneficial to study the extent of human reasoning, to not only inspire the development of artificial machines that reason flexibly and quickly like humans, but to also understand how human reasoning may be impaired in certain circumstances or by certain disorders. To address the extent of human reasoning capability and its applicability to artificial intelligence, we must address three questions: How can humans reason so quickly? What representations or algorithms could lead to reasoning success and failure? And how can these representations and algorithms be realized with neural circuits that could inform the development of intelligent machines?

To tackle the question of reasoning speed, one prominent idea suggests that humans acquire in infancy a set of core knowledge concepts that constrains their expectations of the world and drives their response (Spelke & Kinzler, 2007). Another idea suggests that humans through experience alone learn to form concepts to build a concise world model that could guide their decision making (M. Botvinick et al., 2009; M. Botvinick & Weinstein, 2014; Eckstein & Collins, 2020). These two ideas are not mutually exclusive, as it is possible that a combination of innate knowledge and experience-driven learning is required for the full extent of human reasoning (Kemp & Tenenbaum, 2009). This is the essence of inductive bias, or a set of assumptions other than the data that restricts the learning space and influences the chosen hypothesis in novel circumstances (Baxter, 2000). Inductive biases are characterized as priors in probabilistic models of cognition (Griffiths et al., 2010).

Probabilistic models are sensitive to the exact representation of priors and great success has come from employing logical formulas and programs. For instance, Bayesian program learning exhibited one-shot learning in a handwritten letter task (Lake et al., 2015) and solved novel problems by composing lambda calculus expressions (Ellis et al., 2020). In these frameworks, inductive biases are predefined primitives, also referred to as domain-specific languages, that could be used to construct complex concepts and guide reasoning. These recent successes led some researchers to embrace the long-standing idea that people might employ a similarly discrete and compositional language of thought (Fodor, 1976; Piantadosi, 2021; Quilty-Dunn et al., 2023; Rule et al., 2020). There are strong arguments for having explicit discrete representations to study human reasoning, given that human thoughts are considered compositional, stable and precise (Dietrich & Markman, 2003; Fodor & Pylyshyn, 1988). However, the observation that inductive bias can be represented with discrete programs and symbols does not imply that it has the same representation in the brain.

Inductive bias also found its way into artificial neural networks, most often encoded in the network's architectures to exhibit properties that could constrain the learning process and lead to better generalization (Goyal & Bengio, 2022). For example, convolutional neural networks have translational invariance, meaning they can recognize patterns regardless of their location within the input (Krizhevsky et al., 2017; Lecun et al., 1998). In contrast, graph neural networks (GNNs) encode relational inductive biases, which prioritize the relationships and connections between entities in a dataset (Battaglia et al., 2018). Artificial neural networks with built-in biases have achieved great success in some reasoning tasks (An & Cho, 2020; D. G. T. Barrett et al., 2018; Jahrens & Martinetz, 2020; Kerg et al., 2022; Małkiński & Mańdziuk, 2024; Sinha et al., 2020; Webb et al., 2023). The lack of discrete representations in these neural networks also challenge the language of thought hypothesis. Indeed, the current best models of language itself are deep networks with continuous internal representations (Bubeck et al., 2023). Unfortunately, interpreting the internal representations of neural networks remains a major challenge for the field (Ghorbani et al., 2018), and aligning artificial circuits to the biological circuits employed by humans is no simple feat (D. G. Barrett et al., 2019).

Graph theory might be a potential bridge between discrete representations and neural networks, bringing interpretability of function to connectivity and topology. First and foremost, graphs are good representations for encoding abstraction as well as the relationship between abstract entities, necessary for guiding reasoning (Tenenbaum et al., 2011). Graphs were used in a probabilistic model to encode structural priors and study human learning (Chater et al., 2006; Kemp & Tenenbaum, 2008). Recently, graphs also proved valuable in constraining search in a domain-specific language solution to object-centric reasoning (Y. Xu et al., 2022). On the connectionist side, graph neural networks, useful for learning representations of graphs, are gaining popularity and have achieved astounding performance in algorithmic reasoning tasks such as sorting, searching, dynamic programming, path-finding and geometry (Cappart et al., 2022; Dudzik & Veličković, 2022; Ibarz et al., 2022; K. Xu et al., 2020). Furthermore, besides having relational bias (Battaglia et al., 2018), graph neural networks have many variants implementing other inductive biases in their architectures (Zhang et al., 2022), adding to the expressivity of this tool, and providing promising directions linking architecture and structure to function, especially in the context of reasoning.

Despite growing interest in inductive bias as a cornerstone of human cognition, we still lack a computational framework for systematically testing hypotheses about how such biases are encoded, operate, and shape reasoning. Existing theories often speculate about the role of priors in human learning, yet few offer the means to evaluate these ideas at computational, representational, and mechanistic levels. In this work, we introduce a novel framework that integrates Graph Theory and Graph Neural Networks (GNNs) to model the inductive biases that support human reasoning. Prior studies have either used neural networks over graphs to solve reasoning problems (e.g., Ibarz et al., 2022) or applied graphs to model the structure of human thought (e.g., Tenenbaum et al., 2011), but these approaches have remained largely separate. Our framework bridges this gap. We use graphs to represent structural and abstract priors, GNNs to learn input-output mappings constrained by these priors, and behavioral data to connect model outputs to human-like successes and failures in reasoning.

To evaluate our framework, we applied it as a proof of concept to selected problems from CogARC (Cognitive Abstraction and Reasoning Challenge; Ahn et al., 2024), a behavioral dataset inspired by the Abstraction and Reasoning Corpus (ARC; Chollet, 2019). ARC was designed as an open-ended benchmark for abstract visual reasoning in machines. In this task, solvers are shown a small set of input–output examples and must infer the underlying transformation rules to apply to a novel test input. These rules span diverse visual concepts grounded in core knowledge theory from developmental psychology (Spelke & Kinzler, 2007), making ARC well-suited for studying few-shot generalization.

While recent AI systems have achieved impressive performance on ARC problems (Akyürek et al., 2025; Berman, 2024; Li et al., 2024), much less is known about how humans solve these problems—or why they fail (Acquaviva et al., 2022; Johnson et al., 2021; LeGris et al., 2024; Moskvichev et al., 2023). CogARC addresses this gap by providing 13,606 human trials across 75 ARC problems. In each problem, participants were given a small number of input–output examples and asked to generate an output for a novel test input. Notably, even though each test input has a unique correct answer, participants produced diverse solutions—some correct, others systematic but incorrect—revealing variability in human reasoning.

We hypothesize that these differences arise from distinct inductive biases—prior beliefs about what kinds of entities and relationships matter for solving the problem. We formalize these biases using graphs that encode both structure (e.g., adjacency, connectivity) and abstraction (e.g., what constitutes a node: tile, region, or color). These graph-based priors shape how a solver interprets the task and constrains the space of plausible solutions. Within our framework, we can (1) explicitly define and manipulate these priors, (2) train GNNs to learn ARC input–output mappings under different priors, (3) visualize how computation unfolds under each prior via node/edge importance, and (4) automatically search for priors that reproduce human-like successes and errors.

This framework allows us to reverse-engineer the inductive biases that support or limit generalization. In cases where the GNN receives a well-aligned prior—one that matches the underlying structure and abstraction level of the reasoning problem—it exhibits few-shot generalization, solving the novel test inputs after seeing just a few examples. But when priors are misaligned or incomplete, the model fails in ways that can closely mirror common human errors. These parallels suggest that human reasoning failures may stem not from random noise but from systematic mismatches in prior assumptions. Furthermore, by embedding our graph generation and GNN training loop within a Bayesian optimization pipeline, we can efficiently discover the prior graphs that best explain a given human solution, and identify the structural or abstraction-level components that are essential for success.

Together, these contributions establish a computational bridge between human reasoning and neural network behavior. Our work provides a principled framework for modeling human inductive bias, grounded in graph-based representations and neural computation. By treating priors as modular, testable, and interpretable components, our approach advances our understanding of how reasoning emerges from structured representations—and opens new avenues for building AI systems that reason more like humans.

## Results

**Human Solutions Varied Across Problems. Priors can be Formalized Using Graphs**

Our analysis of human solutions to ARC problems revealed considerable variability in problem-solving strategies. Each ARC problem is defined by a small set of input–output pairs, each illustrating a unique visual transformation rule (**Figure 1A**). Given a novel test input, solvers must infer the underlying rule and generate the appropriate output (**Figure 1B**). Strikingly, even though each problem only has one correct answer, individuals often produce divergent solutions to the same problem (**Figure 1C**). Furthermore, the same solutions are often shared across groups of individuals (Ahn et al, 2024). We hypothesize that solvers rely on distinct priors—preconceived assumptions about structure or abstraction—that shape their interpretation of each problem.

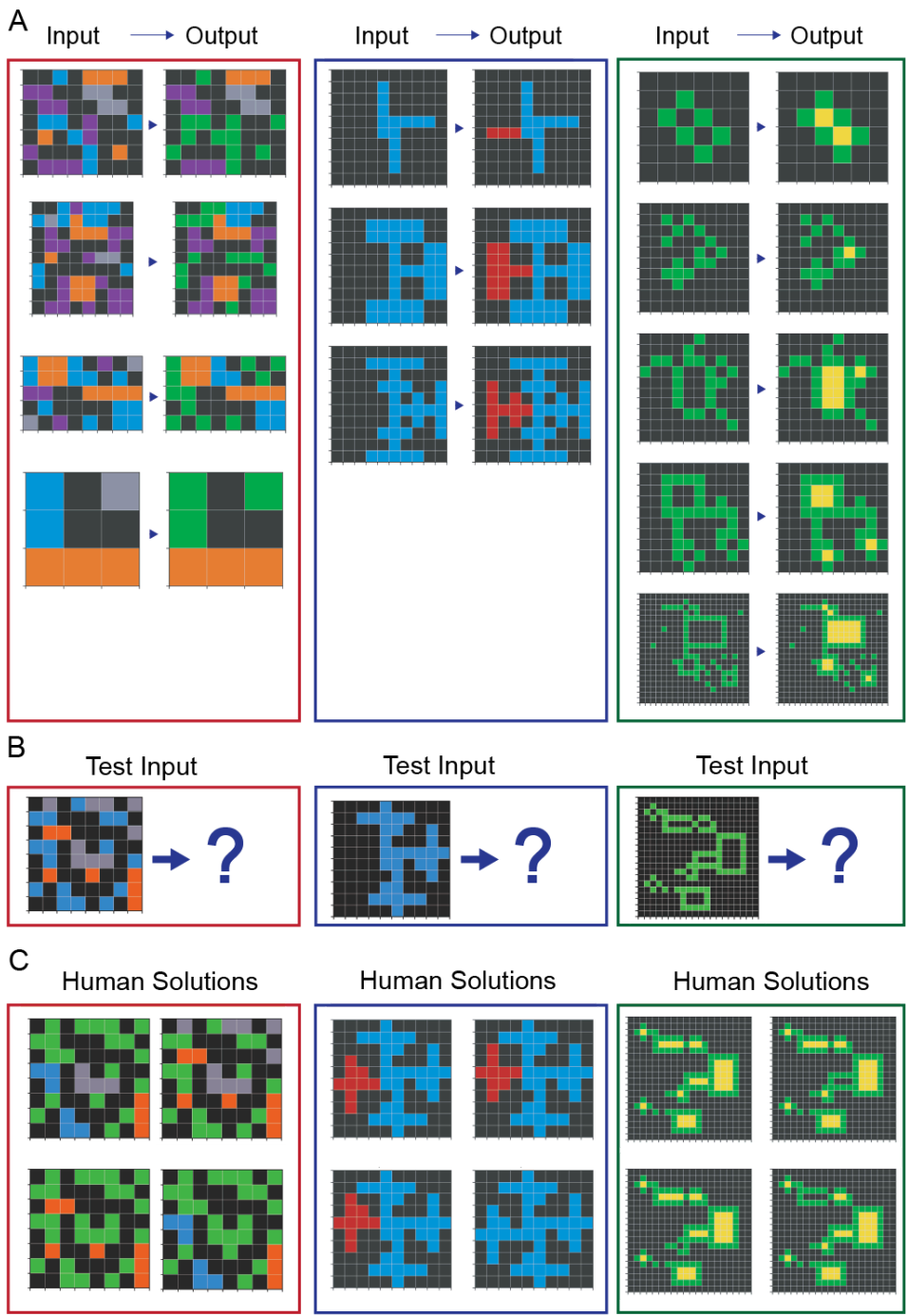

**Figure 1.** *Sample problems from the ARC problem set and human solutions.* **A)** The ARC problem set contains diverse problem types, each defined by a small set of input-output pairs illustrating a unique transformation rule. **B)** Given a novel test input, solvers must infer the underlying rule and generate the correct output. **C)** There is only one correct answer (top left for each panel) but different individuals often produce divergent solutions to the same problem.

These priors, conceptualized as initial hypotheses about which entities and relationships matter, help constrain the space of possible solutions. This facilitates more efficient reasoning and supports few-shot learning. In the ARC setting, such priors guide the solver's interpretation of both the structural relationships between elements and the appropriate level of abstraction (**Figure 2A**).

To formalize these priors, we represent ARC images as graphs (**Figure 2A**). Unlike standard bitmap representations, which treat pixels or tiles independently and lack relational structure (**Figure 2B**), graph representations encode meaningful connections among elements.

In Figure 2C, we vary the *connectivity* of a graph derived from a sample ARC image to impose different structural priors. In all cases, each node corresponds to a single tile in the image and encodes the tile's (x, y) location and color. We compare three variants: (left) a 4-neighbor graph, where edges connect each node to its immediate horizontal and vertical neighbors; (middle) an 8-neighbor graph, which includes diagonal neighbors as well; and (right) a fully connected graph, where every node is connected to every other node. These examples show how different assumptions about which elements are structurally related—local adjacency versus global connectivity—can be encoded through edge patterns.

In Figure 2D, we illustrate how *abstraction* arises by redefining what each node represents, independent of edge structure (i.e., in this example, all graphs are disconnected with no edges). All node representations include (x, y) position and color; node size is added when appropriate. (Left) In the tile-level graph, each node corresponds to an individual tile. (Middle) In the connected-component graph, each node represents a spatially contiguous region of tiles sharing the same color (using 4-neighbor connectivity), including regions of black. Node positions are the centroid of their respective regions, and node sizes reflect the number of tiles in the region. (Right) In the color-grouped graph, all tiles of the same color are aggregated into a single node. Node position is the average (x, y) location of all tiles of that color, and node size reflects the total count of tiles in the group. This third graph abstracts away spatial contiguity, treating color as the main organizing feature. These examples illustrate how abstraction priors determine the level at which information is grouped, and which entities are considered meaningful for reasoning.

Finally, the specific graph selected for a task embodies a solver's prior belief about which entities and relationships are relevant, which could shape both how the problem is interpreted and how a solution is derived (**Figure 2E**).

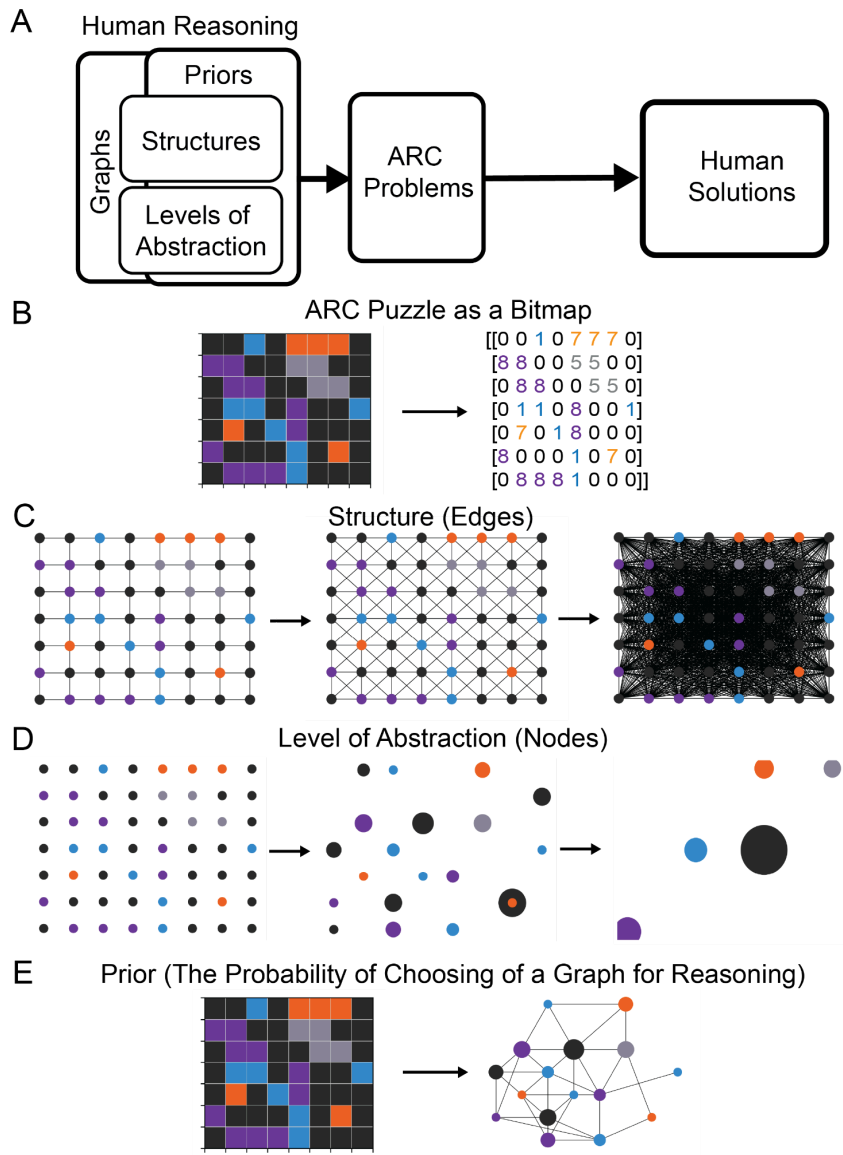

**Figure 2.** *Graph as prior: representing structure and abstraction in the ARC problem set.*
**A)** Human reasoning may rely on priors that encode both structure (relations) and abstraction (entity type), guiding solutions to ARC problems. These priors can be formalized as graphs. **B)** Bitmap inputs treat each tile in isolation and lack relational or abstract structure. **C)** Graphs can impose structure by varying edge connectivity: the same tile-level input is shown with (left) 4-neighbor, (middle) 8-neighbor, and (right) fully connected graphs. These variants determine how nodes relate to their neighbors and shape the inductive prior. **D)** Abstraction is introduced by redefining what a node represents: (left) each node is a tile; (middle) each node is a connected region of same-color tiles (4-neighbor adjacency), placed at the region's centroid and sized by the number of tiles; (right) each node aggregates all tiles of a given color, located at their average position. Node size reflects the number of tiles in the region or color group. **E)** A chosen graph reflects a prior, or specific hypothesis about what entities and relationships matter for reasoning, constraining the space of possible solutions.

## Graph Neural Network with an Appropriate Prior can Exhibit Few Shot Learning and Solve Complex Tasks

We investigated how priors over both graph structure (i.e., connectivity) and level of abstraction (i.e., what each node represents and what features it encodes) influence rule induction and generalization in Graph Neural Networks (GNNs).

We first tested whether simple relational priors—specifically, graphs that encode adjacent-tile connectivity—would be sufficient for generalization on ARC tasks governed by local rules. Using manually constructed graphs where nodes represent individual tiles and edges reflect 8-connected adjacency (**Figure 2C**, middle panel), we trained a GNN on 75 tasks from the CogARC benchmark. The model was trained with only 2 to 6 input–output pairs (**Supplemental Fig. 1A**) and evaluated on a novel test input.

GNNs successfully solved 9 of the 75 tasks using this setup—all involving local rules such as expand, shift, or reflect (**Supplemental Figure 1B,C**). These results show that when the abstraction level (tiles) and structure (local adjacency) are aligned with the underlying rule, the model can induce the correct behavior and generalize from a few examples.

To examine whether variations in abstraction and structure can promote or constrain generalization in a more complex setting, we applied GNNs to a harder ARC task (**Figure 3B, Figure 1A** left panel), where the correct rule involves recoloring connected regions containing adjacent tiles of the same color to green if they contain fewer than three tiles, while ignoring black tiles. Solving this task requires a higher level of abstraction: nodes must represent connected components rather than individual pixels, and the rule depends on region size, not just color or spatial location.

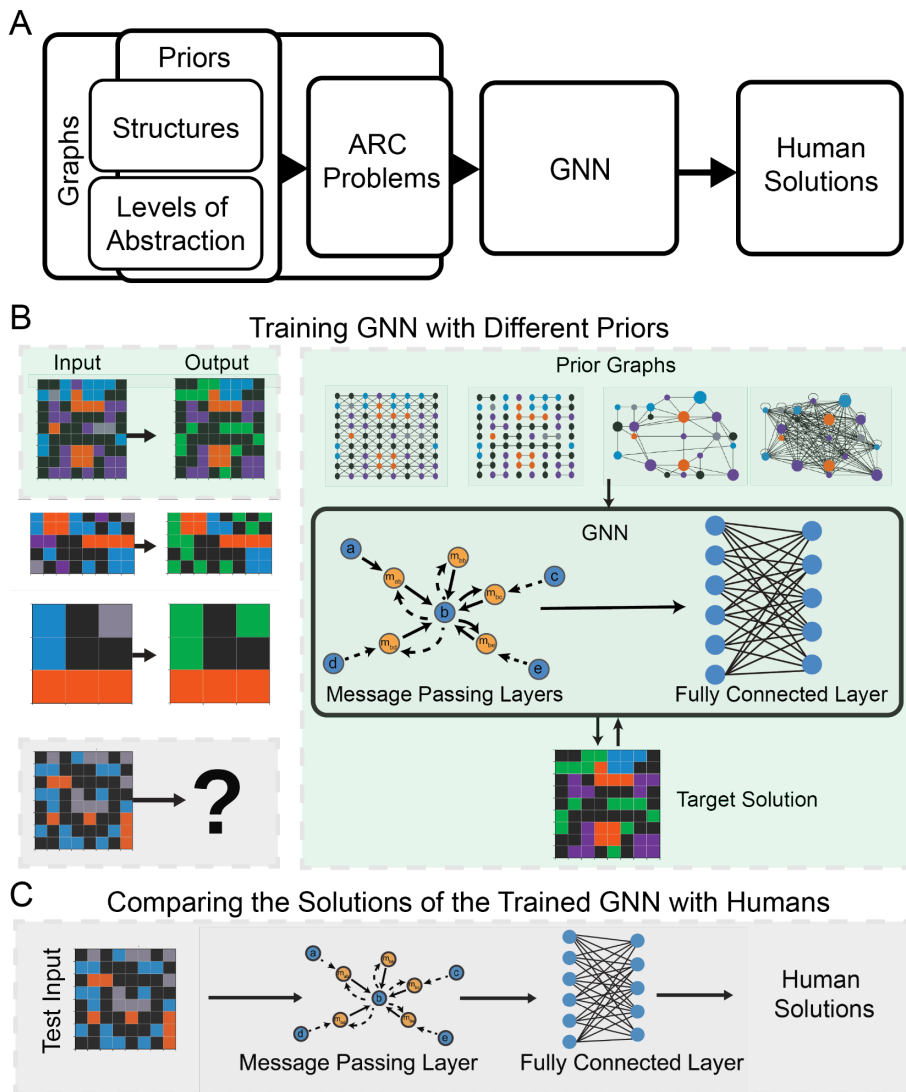

**Figure 3.** *Using a Graph Neural Network (GNN) to map graph-based priors to human solutions.*
**A)** Schematic illustrating how structured priors—capturing relationships and levels of abstraction—can be encoded as graphs and passed to a GNN. The GNN learns to map these graph representations to output solutions. **B)** Example ARC problem (left) with three input–output pairs, in which the rule involves recoloring certain tiles. While human priors are unknown, we can systematically vary the assumed prior by changing the graph representations (top row, 'Prior Graphs'). Each graph reflects a different hypothesis about which entities and relationships are relevant for solving the problem. A GNN is trained on each prior graph using backpropagation. The model consists of message-passing layers followed by a fully connected output layer (middle row). The model is trained only on the input-output example pairs. **C)** After training, the GNN receives a novel test input encoded as a graph and generates a predicted output. This solution can be compared to human responses to assess which priors best support generalization.

We can encode assumptions about abstraction and structure as graph representations, which are then provided to the GNN (**Figure 3A**). Each graph reflects a different hypothesis about what entities exist (e.g., individual tiles vs. connected regions) and how they relate (e.g., fully connected vs. disconnected). By varying these graph-based priors—modifying node definitions and edge connectivity—we can simulate a range of potential assumptions a human solver might bring to the task.

For each prior, we trained the GNN on all of the input–output pairs (3 of which are shown in **Figure 3B**, out of 4 total as shown in **Figure 1A**, left). The GNN consists of message-passing layers followed by a fully connected layer and is trained end-to-end via backpropagation to reproduce the correct outputs for the training examples (**Figure 3B**, right panel, middle row). The message-passing layers enable the model to capture relationships between nodes by aggregating information from their neighbors, reflecting the underlying graph structure. The fully connected layer then takes the learned node representations and maps them to the desired output labels or color transformations, ensuring the abstracted relational information is translated into the correct final predictions.

After training, the GNN was evaluated on a novel test input encoded using the same graph prior. We then compared the model's predicted solution to human responses on the same test input (**Figure 3C**).

In this example problem, human participants showed a range of solutions, including two consistent error patterns: (1) recoloring both blue and gray regions, or (2) recoloring only the blue regions, while consistently leaving the orange region unchanged (**Figure 1C**, left panel). These behavioral patterns suggest that people relied on different priors – in both cases incorrectly assuming the only relevant feature was color, instead of taking into account the number of connected tiles. By observing whether GNNs trained under similarly mismatched priors replicate these errors, we can infer how specific prior assumptions shape generalization performance.

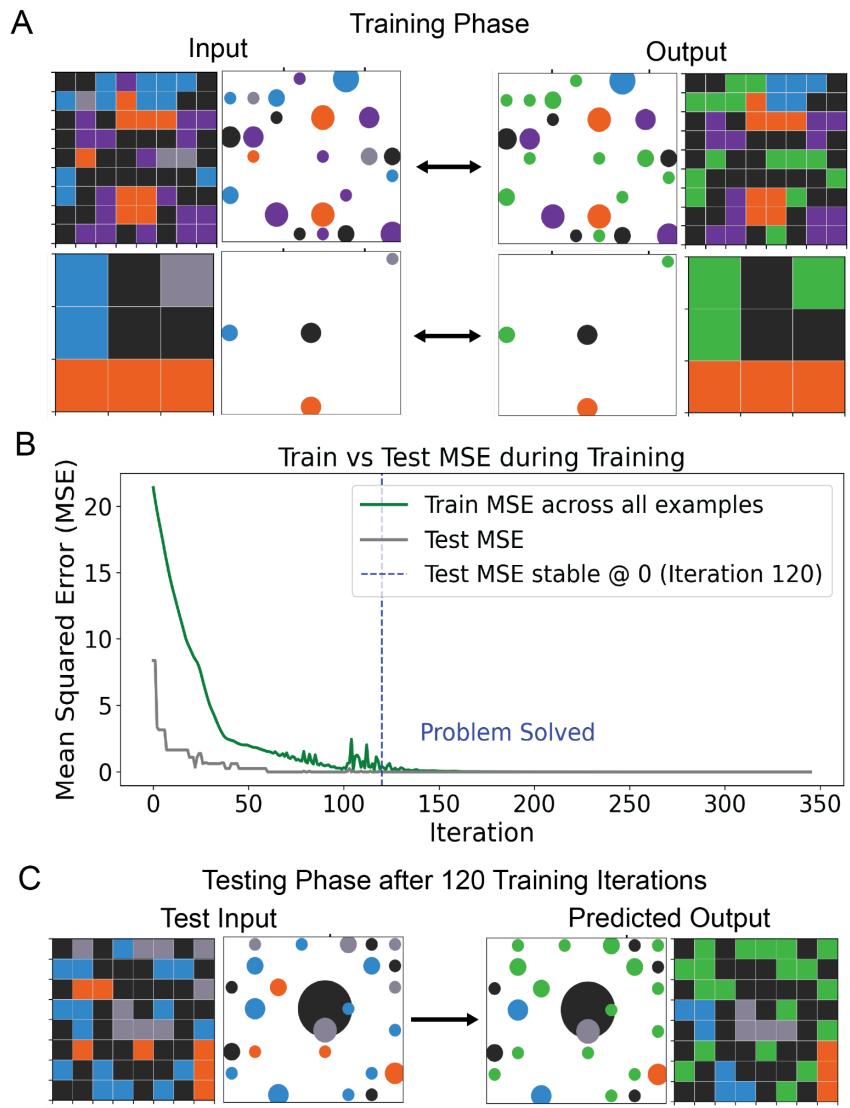

**Figure 4.** *GNNs, when given an appropriate prior, exhibit few-shot learning in a complex problem.* **A)** A graph representation of two ARC input-output pairs from the problem from Figure 3. Each node represents a connected region of uniform color and is characterized by two features: its color and its size (i.e., number of connected tiles). In this example, all edges are removed from the graph. **B)** Training and generalization dynamics of the GNN. The green curve shows the average mean squared error (MSE) between GNN-generated outputs and the correct training outputs. The grey curve shows the test MSE between the GNN's predicted output for the novel test input and the correct solution. The blue dashed line marks when the model solved the problem, defined as the point where test MSE remains at zero for at least 200 consecutive iterations. **C)** Despite limited training examples, the trained GNN generalizes successfully. When the graph priors are as in A) – edges removed and nodes reflecting color and number of connected tiles – the GNN produces the correct solution.

First, to provide the model with the appropriate inductive biases for solving the task, we constructed input graphs where each node corresponded to a connected region of uniform color and carried two features: the region's color and its size (i.e., number of tiles). Crucially, we removed all edges from the graph, since the rule depends solely on counting and does not involve spatial or structural relationships between regions (**Figure 4A**).

Under this abstraction and structure, the GNN was able to solve the task after training on only the 4 examples provided. We evaluated the GNN on the novel test input after each training iteration. We saw that the GNN rapidly converged to zero test error after 120 training iterations on the 4 examples (**Figure 4B**), successfully generalizing to the novel input with different region configurations (**Figure 4C**). The GNN also successfully isolated the black regions and ignored them altogether. This demonstrates that few-shot generalization is possible when the graph prior encodes the correct level of abstraction—connected regions of the same color as entities—and omits irrelevant structural assumptions.

**Human-Like Errors in GNNs Emerge from Mismatched Abstractions and Structural Priors**

We next investigated whether GNNs would exhibit human-like errors when given incorrect or incomplete priors—mirroring those that may underlie failures in human reasoning. To simulate such mismatches, we systematically varied both the abstraction and structure of the prior graphs. As we continued to focus on the same ARC problem, in all cases each node still represents a connected region of uniform color, preserving a basic level of abstraction aligned with the problem. However, to test the source of errors, we removed the size feature from node attributes, depriving the model of a critical cue needed to apply the correct rule. In parallel, we introduced graph connectivity by adding edges between nodes with varying probabilities (**Figure 5A, See Methods**), simulating solvers who impose relational structure between regions—even when such relations are irrelevant to the problem.

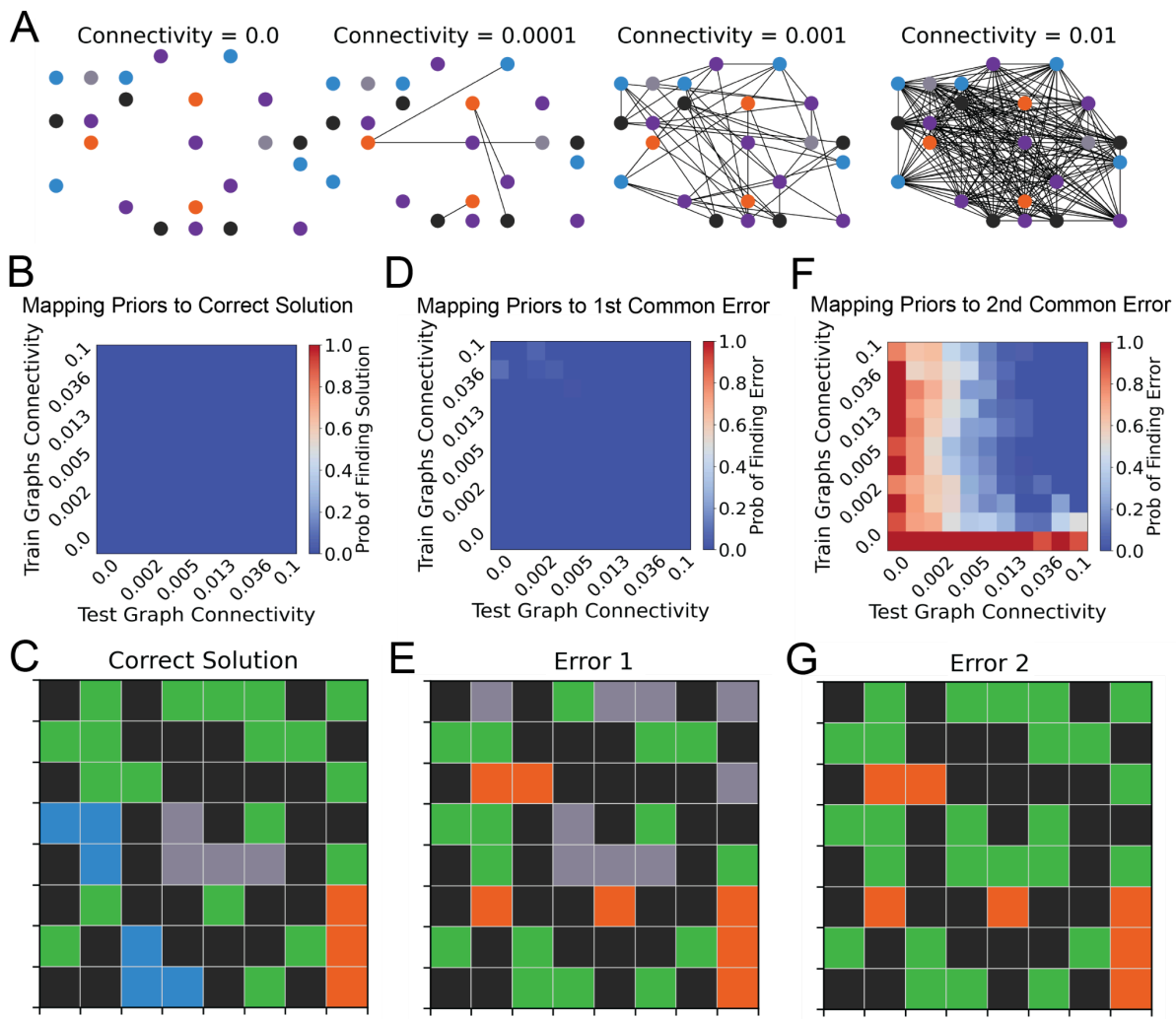

**Figure 5.** *GNNs reproduce human solutions that are sensitive to graph priors.*
**A)** We systematically varied graph connectivity during both training and testing. Node features include only color, while connectivity ranges from 0 (no edges) to 1 (fully connected). Four example graphs illustrate different levels of connectivity: 0, 0.0001, 0.001, and 0.01. **B)** Heatmaps show how successful generalization depends on training and test graph connectivity. The x-axis indicates test graph connectivity; the y-axis indicates connectivity of the training graphs. Color of the heatmap reflects the probability of reproducing a correct solution, with a graded scale from blue (0) to red (1). **C)** The correct solution for the novel test input. **D)** A second heatmap shows the likelihood of finding a specific common human error, conventions as in B. **E)** The most common error made by humans. **F)** A third heatmap shows the likelihood of finding the second most common human error. **G)** The second common error made by humans.

Under these manipulations, the model consistently failed to produce the correct solution (**Figure 5C**). A heatmap of solution accuracy (**Figure 5B**) shows that regardless of the connectivity of the prior graphs, removing size information prevents the model from discovering the correct transformation—highlighting the necessity of encoding region size to solve the problem.

Strikingly, the GNN's failure modes mirrored those observed in human participants. In our behavioral data, we observed two characteristic human error patterns. In one, participants recolored only the blue regions to green (**Figure 5E**); in the other, both blue and gray regions were recolored (**Figure 5G**). In both cases, the orange region was left unchanged, and the black regions were correctly ignored.

To understand when such errors are most likely to occur, we examined model behavior under different structural conditions during both the training and testing phases (**Figure 5D,F**). The probability of producing human-like errors increased when there was a mismatch in connectivity between training and test graphs—for example, when the model was trained on disconnected graphs but tested on connected ones, or vice versa (**Figure 5D,F**). Furthermore, the first error occurred less frequently than the second error (**Figure 5D,F**), indicating that specific combinations of abstraction and structure make certain generalization failures more likely.

Together, these findings show that GNN generalization depends on aligning both the level of abstraction and the structural prior with the rule to be learned. When either is misaligned, the model fails in systematic, human-like ways. Our framework provides a principled approach to studying how reasoning depends on internal assumptions—about what entities exist and how they relate—and reveals how incorrect priors can lead to predictable reasoning errors.

**Optimizing Graph Connectivity Reveals How Structure Shapes Symmetry-based Reasoning**

Instead of manually specifying graph abstractions and connectivity patterns, we developed an automated optimization framework that discovers the structural priors most conducive to human-like generalization. This approach allows us to define a compact set of parameters that govern how graphs are constructed and to search for prior graphs that enable GNNs to both solve the problem with few examples and reproduce common human errors (**Figure 6A**).

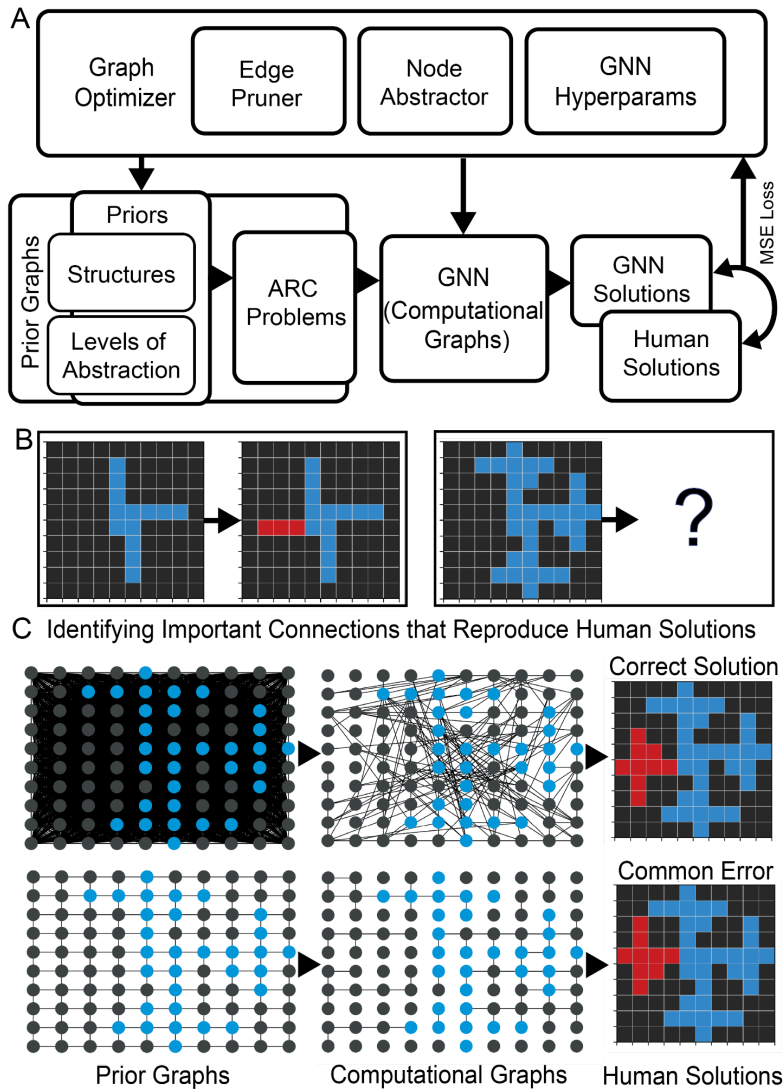

**Figure 6.** *An optimization pipeline for discovering graph priors that reproduce human solutions.*
**A)** Rather than manually designing graphs, we developed a graph optimization pipeline that searches over a parameterized space of graph priors and GNN configurations. The optimizer can adjust node connectivity, abstract or select node features for the prior graphs, and tune GNN hyperparameters. Its goal is to identify the graph structure and model settings that best reproduce a given human solution. **B)** Example ARC problem to illustrate optimization: the rule is to rotate the repeated element to complete the pattern and fill it in red. **C)** The optimizer recovers distinct graph priors and GNN hyperparameters corresponding to different human solutions. Using a leave-one-edge-out strategy, we assess the contribution of each connection in the prior graph, yielding a *computational graph*—a minimal substructure essential for generating the solution that is learned by the GNN during its separate training process. These computational graphs help us interpret how a given prior restricts the learned space of the GNN and supports rule application. In this example, different human solutions (including both correct responses and common errors) emerged from markedly different prior and computational graph configurations.

Our central hypothesis is that a prior graph constrains the space of possible hypotheses, enabling both humans and GNNs to rapidly converge on a particular rule. If this is true, then—when given a human solution—the optimization procedure should be able to recover a prior graph that nudges the GNN toward that same solution.

The optimization process starts from a tile-level fully connected prior graph and applies two key transformations, edge pruning and node clustering. Our optimization pipeline also supports optimizing the GNN's architecture—its width, depth, and learning schedule— but we treat this as optional and defer further details to the discussion section.

To discover which combinations of abstraction and connectivity best align with human first-submission solutions, we embed the full graph-generation and GNN training pipeline within a Tree-structured Parzen Estimator (TPE; Watanabe, 2023) optimization loop. For each CogARC problem, an optimization trial proposes four node clustering parameters, two edge pruning parameters, and three optional GNN's hyperparameters (**See Methods**).

Each inner-loop trial trains a GNN on the input-output example pairs and generates an output prediction from the novel test input. The outer loop evaluates the GNN solution against the human's first-submission solution and minimizes the differences by generating a set of appropriate prior graphs. This yields a set of inductive-bias parameters that best reproduce human reasoning.

But identifying an appropriate prior is only part of the story. To understand what rule the GNN has actually learned under that prior, we developed a method to visualize the computational graph—the subset of nodes and edges critical to the model's prediction. Once optimization selects a prior graph, we systematically remove individual nodes or edges and feed the altered graph to the trained GNN. By comparing the resulting output to the original prediction, we measure how sensitive the model is to each graph element. This leave-one-out strategy highlights the edges and nodes most essential for solving the problem.

We illustrate this process with a problem where the correct rule involves rotating a repeated element and filling it with red (**Figure 6B**, also **Figure 1A**, middle panel). We colloquially refer to this problem as the pinwheel problem. In this problem, we restricted the optimization process to edge pruning, as this alone was sufficient for the optimizer to discover the different graph configurations corresponding to the correct solutions and the common errors.

For the correct solution, the optimization pipeline quickly (within 10 outer loop trials) identified a fully connected prior graph (**Figure 6C**, top left) that provided the connectivity necessary for rotation of the position of the blue tiles. Figure 6C shows only the graph configuration for the novel test input, but the optimizer also found the same fully-connected graph configurations for all the training examples since the graph generating parameters are shared between the training examples and the novel test input (**Supplemental Figure 2A,C,E,G**).
Leave-one-edge-out analysis (see **Methods**) revealed that the long-range diagonal edges were most critical for the GNN's prediction (**Figure 6C**, top middle), aligning with the rule's reliance on rotational symmetry. The GNN uses these diagonal relations to infer the rotated structure and correctly color it red (**Figure 6C**, top right).

For the common human error, where participants mirror the pattern across the vertical axis, the optimizer identifies a sparse 4-connectivity prior graph (**Figure 6C**, second row, left) that connects neighboring pixels along each row and column. This 4-connectivity graph configuration is also shared across all the training examples (**Supplemental Figure 2B,D,F,H**). Computational graph analysis showed that short connecting horizontal edges were used (**Figure 6C**, second row, middle), suggesting the GNN mirrored the pattern by copying elements across the y-axis. This yields the erroneous but plausible mirrored solution seen in many human participants (**Figure 6C**, second row, right).

By pruning edges and visualizing the resulting computational graph, our framework reveals how structural priors constrain the hypotheses a GNN can learn. These priors actively shape the computations that emerge. Our approach offers a transparent, interpretable link between graph structure, learned rules, and human-like reasoning.

**Adding Node Features Enables Discovery of Relational Concepts like Inside–Outside**

In the previous section, the optimizer was limited to pruning edges within a fixed graph, leaving the node definitions untouched. Here, we expand the search space by allowing the optimizer to modify both the structure of the graph (i.e., how nodes are connected) and its abstraction (i.e., how nodes are defined in the first place). Instead of manually deciding how to group tiles or regions into nodes, the optimizer now learns to abstract and connect entities using a combination of UMAP (McInnes et al., 2020) and DBSCAN (Ester et al., 1996). UMAP projects high-dimensional features of the ARC problem into a low-dimensional space, balancing local detail and global structure—trade-offs that the optimizer can adjust. DBSCAN then clusters points in this space based on density, where the optimizer controls whether clusters are tighter or more loosely grouped.

This expanded flexibility enables the discovery of priors that support more complex forms of spatial reasoning. The optimizer tunes three key parameters: the number of nearest neighbors and minimum distance in UMAP, and the clustering radius in DBSCAN. These jointly determine how tile-level features are embedded into a lower-dimensional space, how connectivity is defined in that space, and how regions are grouped into higher-level entities.

Together, these parameters allow the optimizer to generate a wide range of prior graphs that automatically encode both which entities exist and how they relate—a process that previously had to be manually defined.

In particular, the optimized graphs now capture both abstraction and relational structure—distinguishing foreground from background and encoding spatial relationships like surroundedness, which are essential for generalizing on problems that require understanding inside-versus-outside regions.

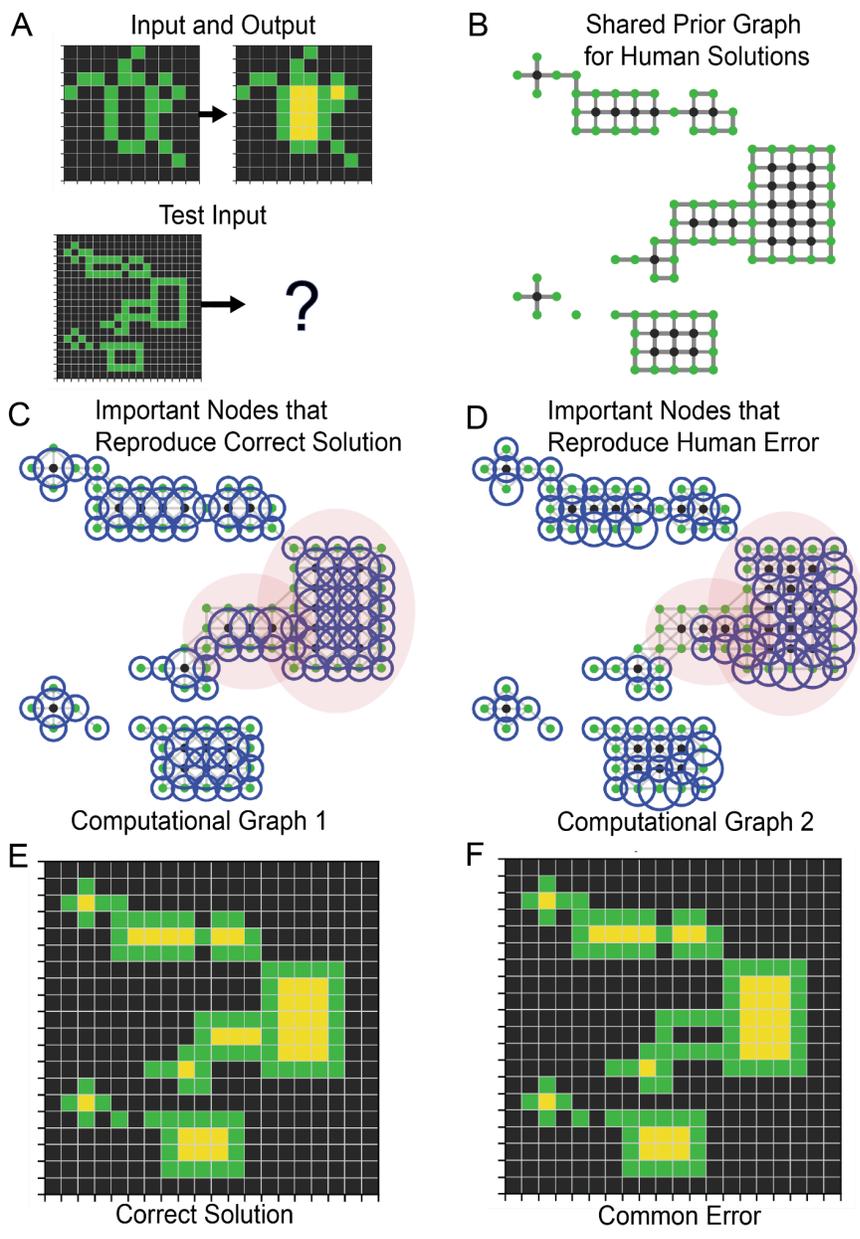

**Figure 7.** *Our optimization framework uncovers inside–outside relations and identifies critical nodes in a graph.* **A)** ARC problem where the rule is to recolor the black region surrounded by green with yellow. Solving this problem requires the notion of background versus foreground, and understanding "inside" and "outside" based on surroundedness. **B)** The optimizer discovered a shared prior graph that reproduces different human solutions. This prior graph contains the black nodes surrounded by a green boundary. **C)** We use a leave-one-node-out strategy to evaluate the contribution of each node to the GNN's prediction. By measuring the impact of removing a node on the output, we identify the *computational graph*—nodes most valued by the GNN for reproducing a human solution. Blue circles highlight the top 100 most important nodes; circle size reflects importance. The red highlighted region indicates a noticeable region of interest between computational graphs (see panel D). This particular computational graph gives rise to the correct solution. **D)** A different computational graph that gives rise to a common human error, despite using the same prior. **E)** Correct solution for this problem. **F)** Common human error in this problem.

We applied this structure- and abstraction-aware optimization to a problem where the goal is to recolor black regions enclosed by green borders with yellow, while leaving all other tiles unchanged (**Figure 1A**, right; **Figure 7A**). Solving this problem depends on recognizing the green boundary as a container and selectively modifying the interior. Strikingly, the optimizer discovered a shared prior graph across multiple human solutions: a graph that includes only the black regions enclosed by green, omitting all the exterior black regions (**Figure 7B**). This prior effectively frames the problem as one of filling the inside, simplifying the learning problem for the GNN. With this abstraction and structure already embedded in the input, the model's job becomes straightforward: recolor all black nodes in the graph to yellow.

Despite using the same prior graph, different GNN solutions emerged—some correct, others erroneous—mirroring the variability seen in human behavior (**Figure 1C**, right). To understand this divergence, we examined what we term the *computational graph*: a subset of nodes deemed most critical to the model's prediction, identified through a leave-one-node-out strategy (see **Methods**). For correct solutions, the GNN relied heavily on the black nodes enclosed by green—precisely those that needed to be recolored (**Figure 7C,E**). In contrast, the common error—where part of the interior was mistakenly left unchanged—was linked to low importance scores for those neglected regions. Although the structural prior was correct, the GNN failed to sufficiently weight the critical nodes during message passing; as a result, these nodes (and their neighbors) did not receive the necessary influence from the learned weights during computation, leading to a failure in updating their colors to yellow.

These findings suggest that human-like errors can arise not only from mismatched priors but also from incomplete or uneven computational focus. Just as people might overlook part of a region, the GNN may underweight critical nodes even when given the correct abstraction. Our framework provides a concrete way to model such oversights, showing how subtle differences in internal computation can lead to divergent outcomes, even under identical input conditions.

Finally, we examined how different training examples contributed to learning. In our setup, the learned GNN's network weights are always shared across all examples, but the prior graph optimizer can work in two ways:

(i) use a single shared set of parameters to generate the prior graphs for all input–output pairs in the problem, which automatically generalize to the novel test input (as illustrated in **Supplementary Figure 2**), or

(ii) fit a distinct set of parameters for each training example (and for the novel test input).

This means that while the GNN learns a common set of weights, the parameters used to generate the prior graphs—which shape the solution space—can, but do not need to be the same across examples. Not every example needs to represent the problem at the same level of abstraction or connectivity. Just a subset of examples with the right abstraction might be enough to push the GNN toward the correct solution, much as we hypothesize may also occur in humans—where a few informative examples could allow a person to infer a rule. Conversely, we hypothesize that a small subset of examples with an incorrect level of abstraction might be enough to push a person toward an error solution.

Thus, using our graph optimizer to fit a distinct set of parameters for each training example, we can assess which training examples yield priors that are helpful or harmful for generalizing to the novel test input (**Supplementary Figure 3**).

In the recoloring-black-to-yellow problem, a helpful prior graph is one that clearly separates the problem-relevant regions—such as isolating the black areas enclosed by green—while excluding irrelevant exterior black nodes. This is the prior graph that the graph optimizer found for the novel test input for both the correct and error solutions (**Figure 7B, Supplementary Figure 3K, L**).

For the correct solution however, by fitting a distinct set of parameters for each training example and for the novel test input, we found even though the test input graph has a clear inside-black-outside-green separation, only two of the five training examples exhibit this relationship (**Supplementary Figure 3A,C**). For the remaining three training examples, the prior graphs only retain green nodes and no black nodes (**Supplementary Figure 3E,G,I**). Even though we are looking only at the input example graphs in this figure, the output example graphs strictly follow the position and color of each node in the input graphs, so without any black nodes, no yellow nodes will be present in the output graphs. The GNN in these training examples will encounter an identity transformation of green-to-green, and these prior graphs will have no impact on the GNN's learned rule of recoloring-black-to-yellow. The first two training examples exhibiting clear inside-black-outside-green separation are sufficient for the GNN to discover and apply the correct recoloring rule (**Supplementary Figure 3A, C**).

For the error solution, only one training example provided this useful abstraction (**Supplementary Figure 3B**). In the rest of the training examples, we found prior graphs that contain a mix of green and black nodes with no clear inside-outside distinction **(Supplementary Figure 3D, F, H, J)**, likely limiting the GNN's ability to generalize.

This highlights how our framework can model the flexibility of human reasoning: some individuals extract the correct rule from just one or two informative examples, while others may fail by overlooking key information across the training set. By capturing how differences in exposure and attention during training shape generalization, our approach provides a more nuanced account of human learning from limited data.

## Discussions

This work introduces a graph-based framework as a tool to bridge abstract reasoning into the domain of cognitive modeling. Our method offers a systematic way to encode structural and abstract priors as graphs, which can be quantitatively manipulated. Our GNN learns input-output mappings, applies them to novel test inputs, and generates solutions that are explicitly shaped by the structure and level of abstraction embedded in the prior graph. This allows us to visualize and analyze how different graph configurations constrain the GNN's computation. Crucially, we show that input–output mappings are not unique; there are many possible transformations that can yield the same output. Our framework enables us to explore this solution space by varying the underlying graph representations. From a cognitive science perspective, our framework provides an entry point into understanding how humans reason. Different types of human errors may correspond to different structural or abstract priors. By exploring and identifying which priors lead to which successes or failures in our model, we begin to map the landscape of reasoning strategies available to humans.

**From Solving Tasks to Understanding Reasoning, a Dialogue Between AI and Cognitive Science via CogARC**

Our goal is not simply to solve problems like ARC, but to use models as instruments to uncover the hidden assumptions—the priors—that humans bring to these problems. By analyzing the human solution, we aim to reverse-engineer the structure and abstraction level that guided the reasoning process. At this early stage, our modelling framework is not meant to be human-like in its reasoning. Rather, it is a framework for testing hypotheses about how humans reason, and for making those hypotheses computationally explicit.

This distinction is crucial in the broader debate between AI engineering and cognitive science. AI researchers often celebrate when a model solves a benchmark task, but cognitive scientists rightly point out that passing the benchmark is not the same as solving it the way a human would (De Cesarei et al., 2021; Lewis & Mitchell, 2024). The cycle continues: a model passes a benchmark, a new benchmark is created, and yet we still lack insight into how humans actually reason. Furthermore, many AI models find shortcuts—they stumble on a solution, but we do not know why or how (Banerjee et al., 2023; Brown et al., 2023; DeGrave et al., 2021). That is not satisfying if the goal is to build models that are not only capable but interpretable and aligned with human cognition.

We use ARC because it captures a broad class of abstract reasoning challenges. Recent large-scale models like GPT have shown impressive performance on ARC (Pfister & Jud, 2025), but again, cognitive scientists argue that these solutions are computationally expensive and do not reflect how humans reason—often efficiently and from very limited data (Mitchell et al., 2023). Furthermore, the response has been to design new ARC-style problems to "break" the models (Chollet et al., 2025). But this reactive loop does not move the field forward. Our framework offers a different path: instead of solving ARC, we use it to understand reasoning. Our long-term goal is to build models that reason as efficiently as humans do, and perhaps one day, surpass human reasoning—but in a way we understand.

**Optimizing Reasoning: Graph Priors, Learning Dynamics, and Extensions to Human-like Computation**

A key technical contribution of our work is the integration of graph optimization with GNN training. The optimizer jointly tunes both the GNN's hyperparameters and the parameters of the prior graphs. For instance, it adjusts the number of message-passing layers, which directly interacts with the range of graph connectivity—whether local or long-range. It also tunes the learning rate, introducing another layer of dynamics that may interact with the structure of the prior graphs. Exploring how graph priors co-evolve with the depth, hidden size, or learning rate of the graph neural network is an important direction for future research.

It is important to clarify the division of roles between the optimizer and the GNN. The optimizer has access to the human solution and uses this to select prior graphs for the example input-output pairs along with the novel test input. Once a set of prior graphs is chosen, the GNN is trained independently—without access to the solution to the novel test input. The GNN must learn to generalize based only on the input-output training examples and the structure and abstraction embedded in the graph. This setup allows us to test whether a learned computation, shaped by the right prior, can succeed on a novel test case. Interestingly, we find that the GNN often doesn't need all training examples to succeed. With just a few well-chosen examples and a meaningful prior, the model can often generalize to the human solution.

This finding raises important modeling questions. Is it always possible to generalize from just a few examples, or do some problems require consistent priors across all training examples? These considerations lead to two complementary strategies. One is to fix the prior graph across all examples, forcing the model to reason with a consistent inductive bias. This approach is especially valuable for modeling individual human reasoning: for example, by identifying a single graph prior that best explains a participant's successes and failures across multiple ARC problems.

The second strategy is to allow the optimizer to select different prior graphs for each example, capturing the diversity of strategies that could support generalization. This opens up a new line of inquiry: how do different training objectives interact with this diversity of priors? For instance, under our current objective, the model is not required to perfectly solve all training examples, mirroring the human tendency to generalize from partial data. But we could impose a stricter learning criterion—minimizing both training and test error—which might pressure the optimizer to discover more coherent priors across examples. This change could let us ask new questions: For instance, is human error correction the result of searching for priors that explain all examples, rather than just one or two? Do people revise their reasoning by minimizing inconsistencies across examples? Our optimization framework offers the tools to test these hypotheses.

Another exciting extension is to model temporal reasoning. At present, our graph representations encode one-step transformations. But many reasoning problems—especially those involving spatial manipulation, sequencing, or trajectory—unfold over time. Humans often solve these problems through sequential steps. For example, in the pinwheel problem, our GNN discovers a shortcut: rather than rotating the pattern, it detects the axis of symmetry and recolors based on that. This is efficient, but may differ from how a human would approach the

problem—perhaps by mentally rotating each element in sequence (Do et al., 2025). Could we build a GNN model that mimics this step-by-step process? Could we visualize the internal temporal dynamics of the network? Extending our framework to model reasoning as a sequence of intermediate transformations is a promising direction for capturing even richer forms of human-like cognition.

**Visualizing Computation and Constraining Priors with Data**

Our results reveal an important ambiguity: multiple priors—each encoding different structures or levels of abstraction—can lead to the same observed solution. This means that human solution alone is not enough to identify the reasoning process that produced it. A model may converge on a correct mapping, but for the wrong reasons, guided by priors that are functionally effective yet cognitively implausible.

This is where our framework offers a key advantage. Because the GNN processes graph-structured inputs, we can visualize how information moves through the network—not just in terms of local or long-range connections, but also in terms of which nodes are grouped together, which ones play a central role, and how those groupings shift depending on the prior. These visualizations help us see which parts of the input the model considers important or functionally related. In doing so, we gain insight into how the model reaches a solution, not just what the solution is. This lets us move beyond surface-level input–output comparisons and ask whether the model's internal computation resembles human reasoning, offering a principled way to compare competing cognitive models.

To push this further, we can rely on additional human data to better constrain the model. Eye-tracking, in particular, offers a detailed look at how people reason through a problem over time. Patterns of fixations and saccades reflect how attention shifts across the visual input—and, more broadly, how people search through different hypotheses (Hollingworth & Bahle, 2020; Liu & van Ede, 2025). By analyzing these eye movements, we can start to infer which parts of the input are being considered, in what order, and at what level of abstraction. This gives us a behavioral window into how priors evolve into concrete computations or rules. We can then compare these human search patterns to how the GNN processes the graph, using them to narrow down the kinds of priors the model should explore. In doing so, we bring the model closer to not just solving the problem, but reasoning in a way that aligns with how humans do it.

**A General Framework to Model Human Reasoning**

So what should a successful framework for tackling ARC look like? A fundamental starting point is to recognize that input–output mappings in ARC are not unique. There are many possible transformations that can turn an input into the correct output. Given enough flexibility, a model can always find some way to solve a problem. But we are not interested in just *any* solution—we are interested in the human-like solution. Which transformation best reflects the inductive biases, abstractions, and structural assumptions that humans use to solve the same problem?

This is what our framework is designed to explore. A model that exhibits human-like reasoning should learn from just a few examples. It should begin drawing a solution in the same step-by-step way a human might, and show systematic patterns of divergence when parameters are perturbed, just as humans show variability across individuals. We want its mistakes to be informative and recoverable in ways that resemble human error correction. We want the internal stages of its computation—how it integrates information, applies rules, and shifts representations—to mirror the human reasoning process.

Ultimately, we want the computational building blocks of the model to reflect how reasoning might occur in the brain. Graphs offer a promising way to do this. Graphs naturally capture structure and abstraction. They provide a flexible substrate for representing both symbolic and spatial relations, and they may offer a bridge between cognitive-level reasoning and neural computation. If reasoning relies on assembling and transforming structured representations, then graphs may be more than a cognitive modeling tool—they may be a good approximation of how neural circuits encode, propagate, and revise knowledge.

This opens a rich space of questions. Can we model abstraction as a dynamic process—where graphs are rewired, compressed, or reorganized in service of generalization? Can we enumerate the set of priors that humans rely on, and build a model that selects among them in context, the way a human would? If so, we may begin to understand why humans reason so quickly in some situations and so slowly in others—why some inferences are immediate and others require careful deliberation.

These are the goals that motivate our framework. ARC gives us a testing ground. But the broader aim is to model intelligence in a way that is data-driven, computationally grounded, and cognitively and neurally plausible.

## Methods
### Behavioral Paradigm: CogARC
We evaluated our framework using CogARC (Cognitive Abstraction and Reasoning Challenge), a few-shot, open-response reasoning task modeled on the Abstraction and Reasoning Corpus (ARC; Chollet, 2019). Seventy-five CogARC problems, each defined by a unique set of visual rules, are presented one at a time via an interactive interface. Participants generate solutions with mouse clicks; their click locations and timings are recorded. Because learned rules do not transfer across problems and each problem can involve multiple, nested conditionals, CogARC resists brute-force guessing and provides rich behavioral traces of individual reasoning strategies.

To compare model and human solutions, we first convert each participant's first-submission response into a bitmap. Although participants may submit up to three attempts, we use only their first submission to capture initial rule induction. Our model will then try to reproduce human solutions.

### Graph Priors: Control of Abstraction and Structure

Every CogARC input bitmap—and its corresponding output—is represented as a pair of graphs. At the finest scale, each tile becomes a node linked to its immediate spatial neighbors. Moving up the abstraction hierarchy, connected tiles of the same color or shape merge into object-level nodes, which can in turn be grouped into higher-order regions based on shared features (e.g., color, size, form). Edges in these graphs encode spatial adjacency and geometric relations at each level.

By systematically varying structure and the level of abstraction, we instantiate a family of "prior graphs," each capturing a distinct representational bias.

Quantitatively, we can define these terms as follows:

- **Structure**: $A \in \{0,1\}^{N \times N}$ : the adjacency matrix of graph G with N nodes
- **Abstraction**: $H \in \mathbb{R}^{N \times F}$ : the node feature matrix of graph G with N nodes and F features per node
- **Prior**: p(G): the probability of choosing graph G in the space of all possible graphs, parameterized by A and H

We can implement and measure how changes in abstraction (cluster size, color homogeneity) and structure (edge connectivity) impact the model's ability to match human first-submission solutions.

**Node Abstraction**

At the lowest level, each tile i in a bitmap is a node carrying the feature vector

$$f_i = [x_i, y_i, c_i, 1]$$

Where $(x_i, y_i)$ are its spatial coordinates, $c_i \in \{0,..., 10\}$ its color label out of 10 color options, and the final "1" encodes unit size. Edges connect spatial neighbors with no initial weighting.

To abstract this raw tile graph into higher-level regions, we apply the following pipeline:

1. **Feature Scaling & Weighting.**
   We min–max scale each dimension of $f_i$ into [0,1] and multiply the color coordinate by a factor $c_w$. This color weighing factor allows us to redistribute the contribution of color to the node clustering process later on. The scaled feature is

$$\tilde{f}_i = scale(f_i)$$

$$\tilde{f}_{i,3} \leftarrow c_w \cdot \tilde{f}_{i,3}$$

2. **UMAP Embedding & Graph Extraction.**
   We embed $\{\tilde{f}_i\}$ into a low-dimensional manifold using UMAP (McInnes et al., 2020) with parameters n_neighbors = $nn$ and min_dist = $d_{min}$. UMAP's internal fuzzy-simplicial graph produces a weighted adjacency matrix A whose nonzero entries $(u, v, A_{uv})$ capture local affinities.
3. **DBSCAN Clustering.**
   We cluster the UMAP embedding with DBSCAN (Ester et al., 1996) (radius ε, min_samples=1), assigning each tile i to a cluster label $l_i \in \{0,..., C - 1\}$. Let
   $$C_j = \{i: l_i = j\}$$
   be the set of tiles in cluster $j$.
4. **Cluster-Level Node Construction.**
   Each surviving cluster j becomes a single node whose attributes are:
   - **Position:** $(\overline{x}_j, \overline{y}_j) = (\frac{1}{|C_j|}\sum_{i \in C_j} x_i, \frac{1}{|C_j|}\sum_{i \in C_j} y_i)$
   - **Dominant Color:** $c_j = mode(\{c_i : i \in C_j\})$
   - **Size:** $s_j = |C_j|$
5. **Edge Weight Aggregation & Pruning.**
   For each pair of clusters (j,k), we compute the average UMAP-graph affinity
   $$w_{ij} = \frac{1}{|\{(u,v) : l_u = j, l_v = k\}|} \sum_{(u,v):l_u=j, l_v=k} A_{uv}$$

We can also add a node deletion step to either randomly or systematically drop a node in the graph, and we can renumber the remaining clusters to 0, … , K - 1.

The result is a **pruned cluster-graph** G whose nodes summarize groups of pixels (with features $(\overline{x}, \overline{y}, c, s)$) and whose edges carry normalized weights w. By varying the four abstraction hyperparameters $(c_w, d_{min}, nn, \varepsilon)$, we quantitatively control:

- **Node count K:** finer abstraction → larger K; coarser → smaller K.
- **Cluster compactness:** via DBSCAN's ε and color-weight $c_w$.
- **Local vs. global connectivity:** via UMAP's number of neighbors $nn$ and min distance $d_{min}$.

n_neighbors controls how many neighbors each node is connected to in the high-dimensional UMAP graph. Smaller values emphasize local detail and produce sparser graphs; larger values prioritize global structure, yielding denser connectivity.

min_dist regulates how tightly UMAP packs points in the embedding, influencing whether nearby points stay distinct or get merged.

ε sets the maximum distance for DBSCAN to consider two points as neighbors, determining the granularity of the resulting clusters.

**Edge Pruning from a Fully Connected Graph**

Independently of how nodes are defined or abstracted, we can modify a graph's adjacency by deleting edges using a multi-step percentile trimming rule. Each node is assumed to carry at least three attributes:

- Position
- Color label
- Size

Edges initially connect all nodes and carry a weight $w_{ij}$, defined by the Euclidean distance between node features (position, color and size). Pruning proceeds over a fixed number of steps, each applying a percentile-based rule to remove weak edges. At step t, we compute the $p_t$-th percentile of the set of edge weights $\{w_{ij}\}$ and remove all edges with $w_{ij} \leq p_t$. The list of percentiles $[p_1, p_2, ..., p_t]$ defines a pruning schedule over T steps, allowing increasingly aggressive removal of edges.

This approach provides a simple implementation and also fine control over the sparsification of the graph's structure.

**Adding Edges from a Null Graph**

We define connectivity as the probability p of adding a new edge between a node and an unconnected node. To introduce these new connections, we loop up to $\binom{N}{2}$ times (i.e., the number of possible undirected edges in a graph with N nodes). In each iteration, we traverse all nodes and, for each node, add a new edge to a randomly selected unconnected node with probability p. No edges are removed during this process. Connectivity is varied by sweeping p over a geometric range from 0.001 to 0.1. Higher values of p lead to denser graphs, as more connections are added over repeated passes.

**Equivariant GNN Architecture and Training**

To process our graph-structured representations, we used an Equivariant Graph Neural Network (EGNN) architecture (Satorras et al., 2022), a message-passing neural network designed to preserve key symmetries in data, particularly equivariance to Euclidean transformations such as translation, rotation, and reflection. This makes EGNNs well-suited for problems where relational reasoning and spatial structure are both relevant, as in the ARC problems.

In standard GNNs, messages are passed between nodes based on features and connectivity, but the position of nodes in space (if used) is typically static or treated as auxiliary. EGNNs go further by updating both node features and node positions, while ensuring that the overall computation is equivariant under SE(3) or E(n) transformations. That is, if the input positions are transformed (e.g., rotated), the output is transformed in the same way, preserving geometric consistency.

Each EGNN layer updates the node features $h_i$ and coordinates $x_i \in \mathbb{R}^d$ (we use d=2 for pixel and spatial data) as follows:

1. Edge messages are computed using both node features and the pairwise distances between node positions:
$$e_{ij} = \phi_e(h_i, h_j, ||x_i - x_j||^2)$$
where $\phi_e$ is a learned neural network (e.g., an MLP).

2. Coordinate updates use the edge messages to nudge the positions of nodes based on their neighbors, in a way that preserves equivariance:
$$\Delta x_i = \sum_{j \in N(i)} (x_i - x_j) \phi_x(e_{ij})$$
where $\phi_x$ is another learned function, and $N_i$ is the neighborhood of node i.

3. Feature updates combine the current features and aggregated messages:
$$h_i^{(l+1)} = \phi_h(h_i^{(l)}, \sum_{j \in N(i)} e_{ij})$$
where $\phi_h$ is a neural network that updates node features using incoming messages.

Importantly, EGNNs do not require explicit encoding of rotation or translation into the model; instead, the equivariance arises naturally from the design of the coordinate and feature update functions.

EGNNs are particularly well-suited to the ARC and CogARC tasks for several reasons:

- They prioritize spatial position, integrating it directly into both feature and message computations.
- They naturally encode relational structure and topological constraints via message passing.
- They can adapt to abstract entities—such as connected regions or color groupings—through flexible node features and graph structure.
- They preserve spatial symmetries, allowing us to examine the effects of priors like connectivity, abstraction, and layout.

In our most basic implementation, graphs can be converted into PyTorch-Geometric `Data` objects and fed into the EGNN. Each node carries its 2D coordinate and one-hot color feature;

edges carry a learned weight. Our EGNN has four layers (hidden size=64), no attention, and is trained with Adam (learning rate = 1e-3, weight decay=1e-5) to predict cluster‑level color labels and, via a masking scheme, per-pixel node labels. We use a combined cross-entropy loss (cluster + node) and early stopping on a held-out validation example. After each epoch, we also compute a test‑set mean squared error (MSE) by reconstructing the full image from cluster predictions.

The full training pipeline is as follows:

1. **Data Preparation**
   - **Node features**: each node's 2D position (x,y) and one-hot encoding of its dominant color (11 channels) form the input tensors x and h.
   - **Edge attributes**: edges carry a single weight feature (after pruning)
   - **Labels**: Node's cluster color label along with the original tile (pixel) color label.
2. **Model Architecture**
   - Four EGNN layers, each with hidden dimension 64 and no attention heads.
   - A final MLP projects to 11 output channels (11 colors).
3. **Loss Functions**
   - **Cluster‑level loss**: cross-entropy between predicted cluster label and true cluster label.
   - **Node‑level loss**: after masking out clusters with label –1, we gather the predicted color for each original tile or pixel and compute cross-entropy against the original pixel color.
   - **Total loss**: the sum of cluster-level and tile-level loss.
4. **Optimization & Early Stopping**
   - **Optimizer**: Adam (learning rate = 1e-3, weight decay = 1e-5).
   - **Training loop**: up to 4000 epochs over the training graphs, summing losses and backpropagating each training batch. Each training batch contains 2 to 6 training examples from ARC.
5. **Test‑Set Evaluation**
   - After every epoch, we reconstruct the solution bitmap from the network's node predictions on the held-out test graph:
     1. Assign each pixel in a kept cluster the cluster's predicted color.
     2. Leave dropped-cluster pixels at their original color.
     3. Compute and report MSE against the human first-submission image.

This training regimen ensures that the GNN learns both a coarse cluster‑level mapping and a fine-grained per-pixel fit, and that its performance—measured by test‑set MSE—directly reflects its ability to reproduce human solutions under only a handful of examples.

**Automated Hyperparameter Search over Graph Priors**
To discover which combinations of abstraction and connectivity best reproduce human first‑submission solutions, we embed the entire graph‑generation and GNN training pipeline within a Tree‑structured Parzen Estimator (TPE) search. Each trial selects:

- For each training input-output pair and the test input:
    - A set of four clustering parameters $d_{min}$, $nn$, $c_w$, $\varepsilon$.
    - A connectivity rule plus its numeric parameters (e.g. threshold, percentile, k).

The inner loop builds the specified prior graphs, converts them into PyTorch‑Geometric `Data` objects, trains our GNN on up to 3–4 input-output pairs (with combined cluster + node cross‑entropy and early stopping), and computes the test MSE against the human's first solution bitmap. The outer loop minimizes the sum of per‑graph training losses and test‑set MSE over up to 10 000 trials, yielding the inductive‑bias configuration most aligned with human reasoning.

The full pipeline is as follows:

### Search Space Definition

- **Abstraction hyperparameters** for each training input-output pair and test input:
    - UMAP $d_{min}$
    - UMAP $nn$
    - DBSCAN $\varepsilon$
    - color-weight $c_w$
- **Structure hyperparameter**: categorical choice of edge modification rule and the associated parameters.
- (Optionally) **GNN hyperparameter:** number of layers, number of units, and learning rates.

### Inner Loop (GNN Training & Evaluation)
For a sampled prior configuration:

- **Graph generation:** build pruned cluster graphs with UMAP and DBSCAN and/or the chosen edge modification rule.
- **Data conversion**: package each pruned clustered graph pair into PyTorch‑Geometric Data objects (node coordinates, one-hot color features, edge index, edge weight, labels).
- **Model fitting:** train our Equivariant GNN on the input-output pairs, using combined cluster + node cross-entropy and early stopping on a held-out validation pair.
- **Validation**: hold out one training example per problem; if its loss fails to improve for 100 consecutive epochs, halt training early.
- **Test performance:** reconstruct the network's predicted solution as a bitmap and compute its MSE against the human first-submission image at each training iteration. Track the lowest MSE seen during training and save the model checkpoint and prior graph whenever the lowest MSE is 0.

**Outer Loop (Bayesian Optimization)**

We use Hyperopt's TPE sampler to minimize the sum of per-graph training losses and the test-set MSE. Each trial proposes full abstraction + structure settings for all problems; after up to 10 000 trials, we recover the graph-prior configuration that leads the EGNN to reproduce human solutions most closely.

**Optimized Configuration**

The recovered configuration specifies for each CogARC problem:

- The optimal clustering granularity $d_{min}$, $nn$, $\varepsilon$, $c_w$.
- The best-performing edge-pruning rule and its parameter(s).
- (Optionally) the GNN's ideal learning rate, depth, and width.

**TPE Sampler Mechanics**

- **Prior initialization**: Hyperopt begins with broad, uninformative priors over each continuous and categorical hyperparameter.
- **Sequential model building**: After an initial set of random trials ("warm-up"), the TPE constructs two density estimates:
  - l(x): the likelihood of hyperparameters x given low observed loss (good trials).
  - g(x): the likelihood of all other trials.
- **Acquisition**: TPE proposes new candidates $x^*$ by sampling where the ratio l(x)/g(x) is highest—i.e., regions of the search space likely to yield improvements.

**Leave-one-edge-out Analysis**

To assess the functional importance of individual edges in the input graph, we employed a leave-one-edge-out (LOEO) procedure, in which each edge is systematically removed and its effect on the model's output is quantified. This approach allows us to estimate the contribution of each edge to the computation performed by a GNN.

The model was pretrained and loaded from a saved checkpoint before evaluation. For all analyses, the GNN was frozen, and no further training occurred during edge importance computation.

Edge Importance Computation

Let G=(V,E) be the input graph with node features and edge connections. The LOEO method proceeds as follows:

1. Baseline Forward Pass: We first pass the full graph through the model and compute the baseline output Y (a prediction for each node's color). This serves as the reference prediction.

2. Edge Removal and Perturbation Testing: For each edge ei ∈ E , we remove ei from the graph and perform multiple forward passes (N=10 by default) through the model using the perturbed graph. Each perturbed graph yields an output prediction X_i.
3. Deviation Scoring: For each run, we compute the mean squared error (MSE) between the baseline Y and the perturbed $X_i$. These scores capture how much the removal of a specific edge alters the model's computation.
4. Averaging and Normalization: The MSE scores for each edge are averaged across runs to produce a single importance score. Finally, all edge scores are min-max normalized to lie in the range [0,1].

Formally, the importance score for edge ei is:

$$Importance(e_i) = \frac{1}{N} \sum_{j=1}^{N} MSE(Y, X_i)$$

To visualize the computed edge importance, we converted the PyTorch Geometric graph into a NetworkX graph and assigned the normalized importance scores to the corresponding edges. Edge weights were visualized using a continuous color scale to highlight edges that most strongly influenced the model's output.

This procedure allows us to uncover which specific relational connections in the graph are most critical for solving a given reasoning task, providing insight into the internal computation performed by the model.

**Leave-one-node-out Analysis**

To assess the importance of individual abstracted units (i.e., clusters of pixels or entities) to the model's output, we performed a leave-one-node-out (LONO) analysis. This procedure systematically removes each node from the graph, quantifies the deviation in model predictions, and identifies the most critical nodes based on the resulting change in model error.

In our framework, graph nodes correspond to abstracted clusters derived from the input image—such as connected regions of the same color or functionally grouped tiles. Each node is associated with a 2D coordinate, a color label, and a cluster identity. These nodes serve as the units over which LONO is computed.

To estimate the contribution of each node (i.e., cluster), we follow a procedure similar in spirit to our leave-one-edge-out analysis:

1. Baseline Output: We compute the model's output on the full input graph G, yielding a prediction for each node (e.g., color logits or class assignments).
2. Node Removal: For each node v, we remove it from the graph (along with its incident edges) and recompute the model output on this perturbed graph. This yields a new prediction for the remaining nodes.

3. Deviation Metric: We compute the difference between the full-graph prediction and the perturbed prediction using a mean squared error (MSE) metric, focusing only on the remaining nodes. The resulting change in error quantifies how much the removal of node v perturbs the computation.
4. Importance Score Assignment: The MSE difference (ΔMSE) serves as the node importance score for v. Nodes with higher ΔMSE values are considered more influential to the final output.

To interpret the learned computation and highlight important components, we visualize the node-level importance scores:

- The graph G is displayed using node coordinates from the original image.
- Each node is colored according to its original ARC color.
- The top-k most important nodes (based on ΔMSE) are overlaid with a thick outline.
- The outline size is scaled proportionally to the normalized importance score, providing a visual indicator of influence.

This LONO analysis provides a computational analogue of lesion studies, allowing us to identify which abstract elements—defined by the model's prior graph structure—are functionally necessary for solving a given reasoning task. By doing so, it enables insight into how the model distributes computation across spatial and conceptual groupings in the input.

**Implementation Details**
All graph processing and visualization are implemented in Python using NetworkX, UMAP, and scikit-learn. GNN training and evaluation use PyTorch and PyTorch Geometric.

**Acknowledgement**

We thank Jingxuan Guo for valuable discussions and advice. This work is supported by the Office of Naval Research ONR MURI N00014-19-1-2571, ONR MURI N00014-16-1-2832, and Kilachand Fund Award.

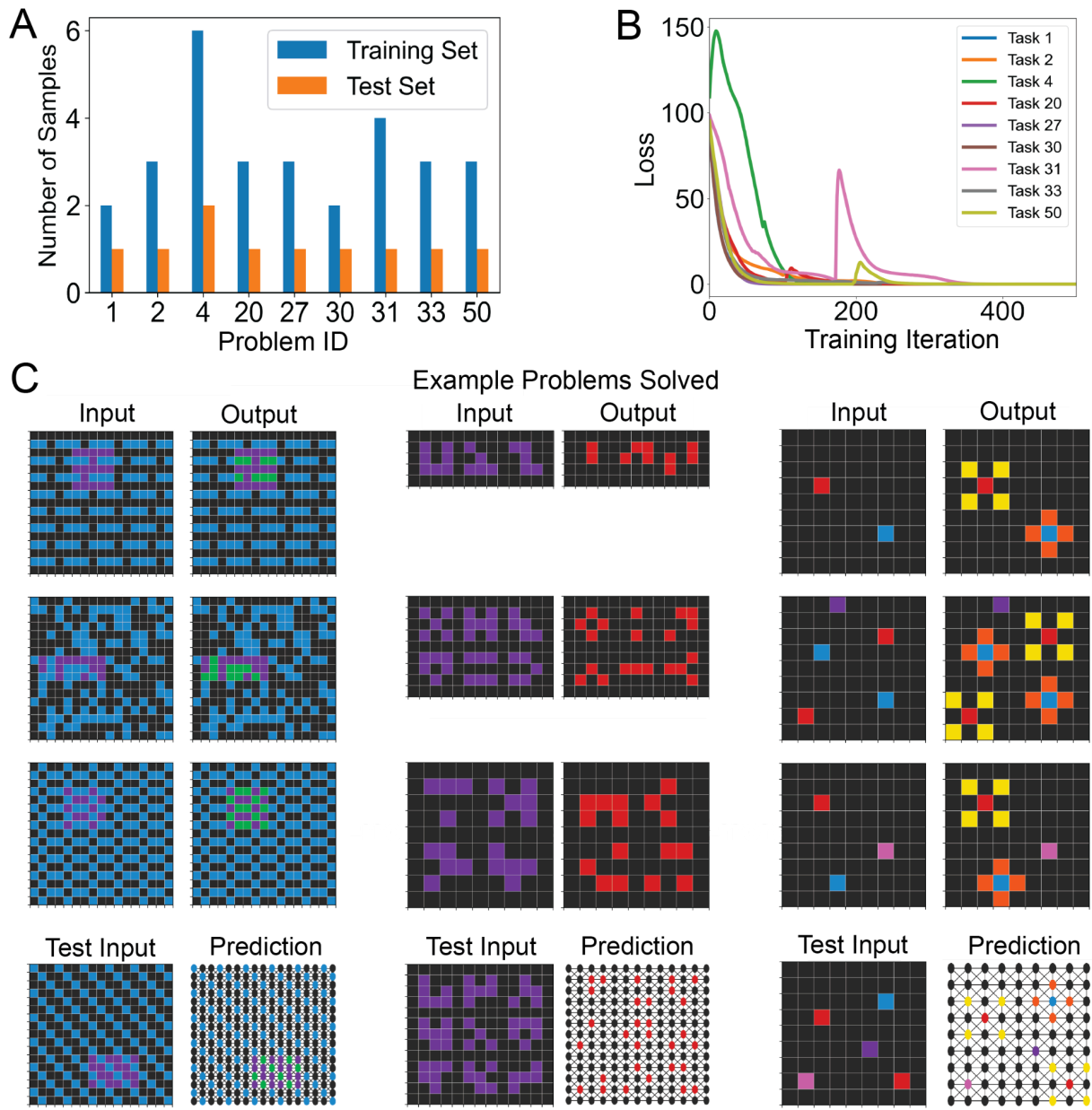

**Supplemental Figure 1.** *GNNs are few-shot learners when given an adjacent-tile graph prior*. A) GNNs are trained with only 2 to 6 examples and tested on up to 2 novel test inputs. B)

Learning curve showing loss over training iteration for the 9 problems GNNs solved when given adjacent-tile graphs as priors. C) 3 example problems solved when GNNs are given adjacent-tile graphs as priors.

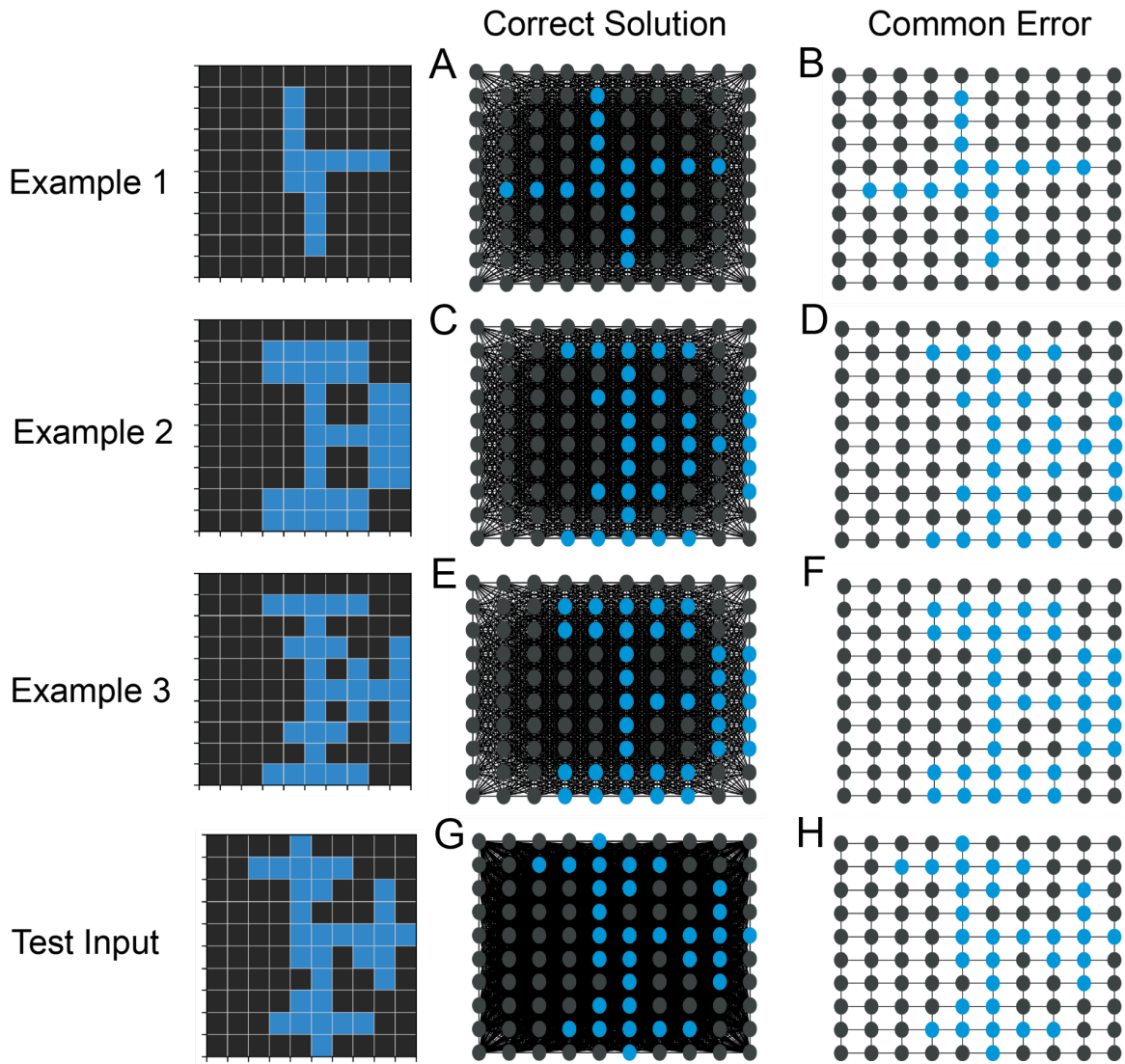

**Supplemental Figure 2.** *Common graph priors for the pinwheel problem.* Using our graph optimizer, we search for common priors — in this case, connectivity patterns — that support a human-like solution for all the training examples. A, C, E, G) Discovered prior graphs for the examples and test input for the correct solution. All are fully connected. B, D, F, H) Discovered prior graphs for the examples and test input for the common error.

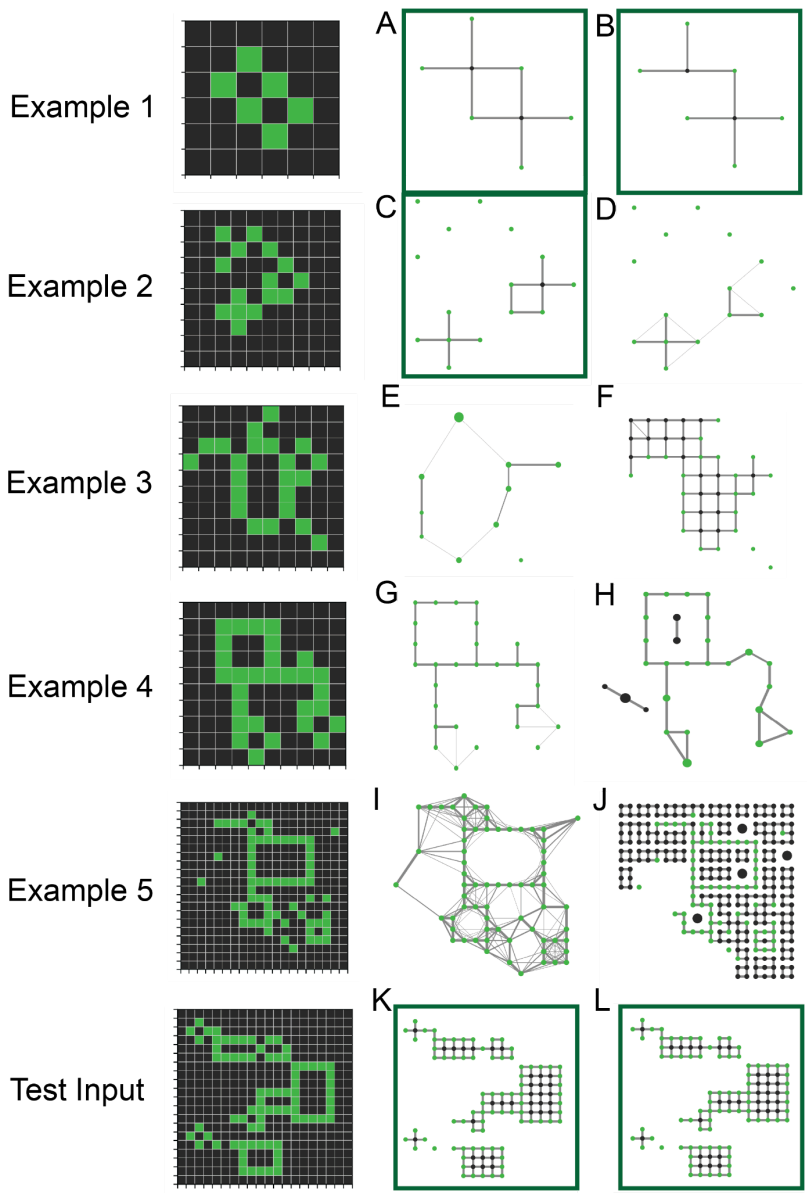

**Supplemental Figure 3.** *Diverging priors for the inside-outside problem.* Using our graph optimizer, we construct and analyze graph priors for each training example—specific node features and connectivity patterns—which allow the model to produce the human solutions for a novel test input. Not all examples require the same representation for successful generalization. In our discovered priors, only two of the five training graphs (highlighted in green) meet the criterion for clear separation—retaining only the green nodes and the black nodes enclosed by green nodes while removing all black exterior or background nodes, regardless of connectivity. Interestingly, the same test graph representation is used by both the correct and error-producing models, indicating that generalization depends on how prior structure is used—not just what it is. A, C, E, G, I, K) Discovered prior graphs for the examples and novel test inputs for the correct solution. B, D, F, H, J, L) Discovered prior graphs for the examples and novel test input for the common error.